\documentclass[prb,twocolumn,showpacs,preprintnumbers,amsmath,floatfix]{revtex4}
\usepackage{amssymb}
\usepackage{graphicx}
\usepackage{bm}
\usepackage{amsmath}
\newcommand{\gl}{{\langle\!\langle}}
\newcommand{\gr}{{\rangle\!\rangle}}

\begin{document}
\title{A fast impurity solver based on equations of motion and decoupling}
\author{Qingguo Feng}
\affiliation{Institut f\"ur Theoretische Physik,
Goethe-Universit\"at Frankfurt, Max-von-Laue-Stra{\ss}e 1, 60438
Frankfurt am Main, Germany}
\author{Yu-Zhong Zhang}
\affiliation{Institut f\"ur Theoretische Physik,
Goethe-Universit\"at Frankfurt, Max-von-Laue-Stra{\ss}e 1, 60438
Frankfurt am Main, Germany}
\author{Harald O. Jeschke}
\affiliation{Institut f\"ur Theoretische Physik,
Goethe-Universit\"at Frankfurt, Max-von-Laue-Stra{\ss}e 1, 60438
Frankfurt am Main, Germany}

\date{\today}

\begin{abstract}
In this paper a fast impurity solver is proposed for dynamical mean
field theory (DMFT) based on a decoupling of the equations of motion
for the impurity Greens function. The resulting integral equations are
solved efficiently with a method based on genetic algorithms. The
Hubbard and periodic Anderson models are studied with this impurity
solver. The method describes the Mott metal insulator transition and
works for a large range of parameters at finite temperature on the
real frequency axis.  This makes it useful for the exploration of real
materials in the framework of LDA+DMFT.
\end{abstract}
\maketitle

\bigskip

\section{Introduction}

Understanding exotic physical properties, such as high-T$_c$
superconductivity and the correlation-driven Mott metal-insulator
transition of strongly correlated compounds (typically those including
$d$ or $f$ electrons), remains a hard and fundamental task in modern
condensed matter physics. During the past decade, the development and
application of dynamical mean-field theory (DMFT) has led to a
considerable improvement in our understanding of these
systems.~\cite{georges96,metzner89,georges92}. The essence of DMFT is
to map a many-electron system to a single impurity atom embedded in a
self-consistently determined effective medium by neglecting all the
spatial fluctuations of the self-energy. However, this resulting
quantum impurity model remains a fully interacting many-body problem
that has to be solved, and the success of DMFT depends on the
availability of reliable methods for calculation of the local
self-energy of the impurity model.

Accordingly, much effort has been devoted to develop various
impurity solvers. Among those, the iterated perturbation theory
(IPT)~\cite{georges92,Zhang,Kajueter,Fujiwara}, the non-crossing
approximation (NCA)~\cite{Keiter,Grewe,Kuramoto,Haule},
equation-of-motion (EOM) method~\cite{jeschke05,zhu05,qi08}, Hubbard
I approximation (HIA)~\cite{Hubbard}, fluctuation exchange (FLEX)
approximation~\cite{Chioncel,Drchal}, the quantum Monte Carlo method
(Hirsch-Fye algorithm)
(HF-QMC)~\cite{Hirsch,Jarrell,Georges,Rozenberg}, the continuous
time quantum Monte Carlo method (CTQMC)~\cite{Werner,Rubtsov}, the
exact diagonalization (ED)~\cite{Caffarel,Si}, the numerical
renormalization group (NRG) method~\cite{Bulla1,Bulla2}, and the
density matrix renormalization group (DMRG)
method~\cite{Garcia,karski08} are widely adopted. However, every
impurity solver has its own limitation. IPT originally cannot be
applied to the case away from half-filling while a modified IPT
which can solve this problem has to introduce an {\it ansatz} to
interpolate the weak and strong coupling limits, and the
generalization of IPT to the multi-orbital case requires more
assumptions and approximations. NCA cannot yield the Fermi liquid
behavior at low energies and in the low temperature limit. The HIA
can only be applied to strongly localized electron systems like $f$
electrons.  FLEX works well in the metallic region while it fails in
the large $U$ region. Before the appearance of CTQMC, the HF-QMC was
not applicable in the low temperature limit and has serious
difficulties in application to multi-orbital systems with spin-flip
and pair-hopping terms of the exchange interaction since the
Hubbard-Stratonovich transformation~\cite{Held} cannot be performed
in these systems. But even for CTQMC, the requirement to do
analytical continuation of the results to the real frequency axis
remains, which introduces some uncertainty especially for
multi-orbital systems. In the ED method, an additional procedure is
required for the discretization of the bath and as a consequence,
the method is unable to resolve low-energy features at the Fermi
level. NRG aims at a very precise description of the low-frequency
quasiparticle peaks associated with low-energy excitations while it
has less precision in the Hubbard bands which are important in
calculating the optical conductivity. Furthermore, all the
numerically exact impurity solvers QMC, ED, NRG and DMRG are
computationally expensive.

However, today, a fast and reliable impurity solver is really urgently
needed due to the fact that great achievements have been made in
understanding correctly the strongly correlated systems from first
principle by combining DMFT and local density approximation (LDA) in
density functional theory (DFT), so called
LDA+DMFT~\cite{Kotliar}. The aim of this paper is to present a fast
and reliable impurity solver based on the EOM method. Equation of
motion methods are limited by their decoupling scheme, but EOM has
shown its value by working directly on the real frequency axis and at
very low temperature. It can be a good candidate for a faster and
reliable impurity solver by choosing a suitable decoupling scheme.  In
fact, the infinite $U$ case was studied by the EOM method for Hubbard
model, periodic Anderson model and $pd$ model in
Ref.~\onlinecite{jeschke05}. In Ref.~\onlinecite{zhu05}, the finite
$U$ case is studied without calculating the physical quantities
self-consistently. Recently~\cite{qi08}, this method has been improved
by taking into account selfconsistency and applied to the Anderson
impurity model in the large-$N$ limit.  The operator projection method
(OPM)~\cite{Onoda01} is related to the EOM method.  In this
paper, we will use a different decoupling procedure than used
previously for a set of EOMs of the Anderson impurity model and then
apply this new impurity solver to the finite $U$ Hubbard model as well
as periodic Anderson model via dynamical mean field theory. Meanwhile,
we employ genetic algorithms to efficiently search for the
self-consistent solution. The genetic algorithm significantly reduces
the CPU time for convergence and improves the energy resolution in the
DMFT calculation.

The paper is organized as follows: In Section~\ref{sec:eom} we
present the EOMs we use and introduce our decoupling scheme. In
Section~\ref{sec:ga} we describe how the genetic algorithm is
implemented in our DMFT loop. Finally, in Section~\ref{sec:results} we
test our EOM impurity solver on the Hubbard model and periodic
Anderson model.

\section{Equations of motion and decoupling procedure}
\label{sec:eom}

We start with the Hamiltonian of the single impurity Anderson model. For
arbitrary degeneracy $N$, it is given by
\begin{eqnarray}
{\cal
H}&=&\sum_{k\sigma}\varepsilon_{k}c^{\dag}_{k\sigma}c_{k\sigma}+\sum_{\sigma}\varepsilon_f
f^{\dag}_{\sigma}f_{\sigma}+\frac{U}{2}\sum_{\sigma\sigma'}\hat{n}_{\sigma}\hat{n}_{\sigma'}\nonumber\\
&&\qquad\qquad+\sum_{k\sigma}\big(V^{\ast}_{k\sigma}c^{\dag}_{k\sigma}f_{\sigma}+V_{k\sigma}f^{\dag}_{\sigma}c_{k\sigma}\big)\label{eq:SIAMhamiltonian}
\end{eqnarray}
where $c^{\dag}_{k\sigma}$, $c_{k\sigma}$, $f^{\dag}_{\sigma}$ and
$f_{\sigma}$ are the creation and annihilation operators for the
conduction electrons and for the correlated impurity electrons,
respectively. $\hat{n}_{\sigma}=f^{\dag}_{\sigma}f_{\sigma}$
corresponds to the density of the $f$ electrons. $\varepsilon_k$ is
the dispersion of the conduction electrons, $\varepsilon_f$ is the
site energy of the correlated electron, $U$ is the on-site Coulomb
interaction strength of the $f$ electrons, and $V_{k\sigma}$ is hybridization
between conduction and correlated electrons.

In studying the system described by the Hamiltonian of
Eq.~\ref{eq:SIAMhamiltonian}, we consider the double time
temperature-dependent retarded Greens function in Zubarev notation~\cite{zubarev60},
\begin{eqnarray}
G_{AB}(t,t')=\gl A(t);B(t')\gr=-i\Theta(t-t')\langle
[A(t),B(t')]_+\rangle
\end{eqnarray}
involving the two Heisenberg operators $A(t)$ and $B(t')$. It is
convenient to work with the Fourier transform, which is defined as
\begin{eqnarray}
\gl A;B\gr_{\omega}=\int^{\infty}_{-\infty}dt e^{i\omega(t-t')}\gl
A(t);B(t')\gr\,.
\end{eqnarray}
In the framework of the equation of motion method, the Greens
function should satisfy the equation of motion
\begin{eqnarray}
\omega\gl A;B\gr=\langle[A,B]_+\rangle+\gl[A,{\cal H}];B\gr
\end{eqnarray}
where we have neglected the lower indices $\omega$. In the following,
all the Greens functions depend on frequency $\omega$.

As a result of the coupling between conduction and $f$ electrons, we find
the equations of motion,
\begin{widetext}
\begin{eqnarray}
(\omega-\varepsilon_d-\Delta)\gl f_{\sigma};f^{\dag}_{\sigma}\gr
&=&1+(N-1)U\gl
\hat{n}_{\sigma'}f_{\sigma};f^{\dag}_{\sigma}\gr_{\sigma\neq\sigma'}\label{eq:nff}\\
(\omega-\varepsilon_d-U)\gl
\hat{n}_{\sigma'}f_{\sigma};f^{\dag}_{\sigma}\gr
&=&\bar{n}_{\sigma'}+(N-2)U\gl \hat{n}_{\sigma''}\hat{n}_{\sigma'}f_{\sigma};f^{\dag}_{\sigma}\gr+\sum_k(-V^*_{k\sigma'}\gl c^{\dag}_{k\sigma'}f_{\sigma'}f_{\sigma};f^{\dag}_{\sigma}\gr\nonumber\\
& &+V_{k\sigma}\gl
\hat{n}_{\sigma'}c_{k\sigma};f^{\dag}_{\sigma}\gr+V_{k\sigma'}\gl
f^{\dag}_{\sigma'}c_{k\sigma'}f_{\sigma};f^{\dag}_{\sigma}\gr)\\
(\omega-\varepsilon_{k})\gl
\hat{n}_{\sigma'}c_{k\sigma};f^{\dag}_{\sigma}\gr
&=&V^*_{k\sigma}\gl
\hat{n}_{\sigma'}f_{\sigma};f^{\dag}_{\sigma}\gr+\sum_{k'}(-V^*_{k'\sigma'}\gl
c^{\dag}_{k'\sigma'}f_{\sigma'}c_{k\sigma};f^{\dag}_{\sigma}\gr+V_{k'\sigma'}\gl
f^{\dag}_{\sigma'}c_{k'\sigma'}c_{k\sigma};f^{\dag}_{\sigma}\gr)\\
(\omega-\varepsilon_{k})\gl
f^{\dag}_{\sigma'}c_{k\sigma'}f_{\sigma};f^{\dag}_{\sigma}\gr
&=&\langle f^{\dag}_{\sigma'}c_{k\sigma'}\rangle+V^*_{k\sigma'}\gl
\hat{n}_{\sigma'}f_{\sigma};f^{\dag}_{\sigma}\gr+\sum_{k'}(-V^*_{k'\sigma'}\gl
c^{\dag}_{k'\sigma'}c_{k\sigma'}f_{\sigma};f^{\dag}_{\sigma}\gr\nonumber\\
&&+V_{k'\sigma}\gl
f^{\dag}_{\sigma'}c_{k\sigma'}c_{k'\sigma};f^{\dag}_{\sigma}\gr)\\
(\omega+\varepsilon_{k}-2\varepsilon_{d}-U)\gl
c^{\dag}_{k\sigma'}f_{\sigma'}f_{\sigma};f^{\dag}_{\sigma}\gr
&=&\langle c^{\dag}_{k\sigma'}f_{\sigma'}\rangle+2(N-2)U\gl c^{\dag}_{k\sigma'}\hat{n}_{\zeta}f_{\sigma'}f_{\sigma};f^{\dag}_{\sigma}\gr_{\substack{\zeta\neq\sigma\\ \zeta\neq\sigma'}}-V_{k\sigma'}\gl
f^{\dag}_{\sigma'}f_{\sigma'}f_{\sigma};f^{\dag}_{\sigma}\gr\nonumber\\
& &+\sum_{k'}(V_{k'\sigma}\gl
c^{\dag}_{k\sigma'}f_{\sigma'}c_{k'\sigma};f^{\dag}_{\sigma}\gr+V_{k'\sigma'}\gl
c^{\dag}_{k\sigma'}c_{k'\sigma'}f_{\sigma};f^{\dag}_{\sigma}\gr)\label{eq:cfff}
\end{eqnarray}
\end{widetext}
where
$\Delta(\omega)=\sum_k\frac{V^{\ast}_{k\sigma}V_{k\sigma}}{\omega-\varepsilon_k}$
is the hybridization function and we have used
\begin{eqnarray}(\omega-\varepsilon_k)\gl
c_{k\sigma};f^{\dag}_{\sigma}\gr=V^{\ast}_{k\sigma}\gl
f_{\sigma};f^{\dag}_{\sigma}\gr\,.
\end{eqnarray}
These equations are generalized to arbitrary degeneracy $N$ compared to
Ref.~\onlinecite{lacroix81}, {\it i.e.} they are at the same level as
Ref.~\onlinecite{czycholl85}. Now a decoupling scheme is needed to
truncate the equations of motion in order to get a closed set of
equations. Here we have used the cluster expansion scheme proposed in
Ref.~\onlinecite{wang94} where the higher order Greens functions are
separated into connected Greens functions of the same order and lower
order Greens functions.  The connected Greens function can not be
decoupled any further as defined. This expansion scheme gives a
natural and systematical way for truncation. It has been used in
Ref.~\onlinecite{luo99} for studying the single impurity Anderson
model, in particular for infinite interaction strength $U$. This
approach to decoupling could be used to study the EOM method beyond
the level of Ref.~\onlinecite{lacroix82}. The detailed cluster
expansion scheme is given as
\begin{eqnarray}
\gl 1;2\gr&=&\gl 1;2\gr_c\\
\gl 123;4\gr&=&\gl
123;4\gr_c+{\hat{A}s}_{(2,3)}\langle 12\rangle\gl
3;4\gr\\
\gl 12345;6\gr&=&\gl
12345;6\gr_c\nonumber\\
&+&{\hat{A}s}_{(2,4,5)}{\hat{S}p}_{(1,2;3,4)}\big(\langle
12\rangle\langle 34\rangle\gl 5;6\gr\nonumber\\
&+&\langle 1234\rangle_c\gl 5;6\gr+\langle 12\rangle\gl
345;6\gr_c\big)~~~~~
\end{eqnarray}
where digits 1-6 stand for operators, $\hat{A}s_{(i,j,k)}$ is the
antisymmetrization operator for operators $(ijk)$,
$\hat{S}p_{(1,2;3,4)}$ is the symmetrization operator for pair
exchange between $(1,2)$ and $(3,4)$, and Greens functions or
correlations marked by an index $c$ represent connected terms.

Using this decoupling scheme and neglecting all the three-particle
connected Greens functions and those two-particle connected Greens
functions which involve two $c$ operators, {\it i.e.} $\gl
c^+_{k\sigma'}c_{k'\sigma'}f_{\sigma};f^{\dag}_{\sigma}\gr_c$, $\gl
c^+_{k\sigma'}f_{\sigma'}c_{k'\sigma};f^{\dag}_{\sigma}\gr_c$ etc., and
by assuming correlations with spin flip to be zero, {\it e.g.}
$\langle f^{\dag}_{\sigma'}f_{\sigma}\rangle=0$, we can get the single electron
Greens function for arbitrary degeneracy $N$ at the level of
approximation of Ref.~\onlinecite{lacroix82} as
\begin{eqnarray}
(\omega-\varepsilon_d-\Delta-AB)\gl
f_{\sigma};f^{\dag}_{\sigma}\gr=1+A\big\{\bar{n}_{\sigma'}+C\big\}\label{eq:arbitraryNGF}
\end{eqnarray}
where
\begin{eqnarray}
A&=&\frac{(N-1)U}{\omega-\varepsilon_d-U-(N-2)U\bar{n}_{\sigma''}-2\Delta-\tilde{\Delta}}\label{eq:arbitraryNA}\\
B&=&\bigg[(N-2)U\langle\hat{n}_{\sigma''}\hat{n}_{\sigma'}\rangle_c+\sum_{k,k'}\Big(-\frac{V_{k\sigma'}V^*_{k'\sigma'}\langle c^+_{k'\sigma'}c_{k\sigma'}\rangle}{\omega-\varepsilon_k}\nonumber\\
&&+\frac{V_{k\sigma'}V_{k'\sigma}V^*_{k'\sigma}}{(\omega-\varepsilon_k)(\omega-\varepsilon_{k'})}\langle f^{\dag}_{\sigma'}c_{k\sigma'}\rangle\Big)\nonumber\\
&-&\sum_k\frac{2(N-2)U\langle\hat{n}_{\sigma''}c^+_{k\sigma'}f_{\sigma'}\rangle_c}{\omega+\varepsilon_k-2\varepsilon_d-U-2(N-2)U\bar{n}_{\sigma''}}\nonumber\\
&-&\sum_{k}\frac{V^*_{k\sigma'}\sum_{k'}(\frac{V_{k'\sigma}V^*_{k'\sigma}}{\omega-\varepsilon_{k'}}\langle
c^+_{k\sigma'}f_{\sigma'}\rangle+V_{k'\sigma'}\langle
c^+_{k\sigma'}c_{k'\sigma'}\rangle)}{\omega+\varepsilon_k-2\varepsilon_d-U-2(N-2)U\bar{n}_{\sigma''}}\bigg]~~~~~
\label{eq:arbitraryNB}\end{eqnarray}
\begin{eqnarray}
C&=&\sum_k\Big(\frac{V_{k\sigma'}\langle
f^{\dag}_{\sigma'}c_{k\sigma'}\rangle}{\omega-\varepsilon_k}\nonumber\\
&&-\frac{V^*_{k\sigma'}\langle
c^+_{k\sigma'}f_{\sigma'}\rangle}{\omega+\varepsilon_k-2\varepsilon_d-U-2(N-2)U\bar{n}_{\sigma''}}\Big)~~~~~\label{eq:arbitraryNC}\\
\tilde{\Delta}&=&\sum_k\frac{V^*_{k\sigma'}V_{k\sigma'}}{\omega+\varepsilon_k-2\varepsilon_d-U-2(N-2)U\bar{n}_{\sigma''}}
\label{eq:Deltatilde}\end{eqnarray}
in which $\bar{n}_{\sigma'}=\langle \hat{n}_{\sigma'}\rangle$. This set of equations \eqref{eq:arbitraryNGF}-\eqref{eq:Deltatilde} is closed by the following two equations for the two-particle connected correlations:
\begin{widetext}
\begin{equation}\begin{split}
&\langle \hat{n}_{\sigma''}\hat{n}_{\sigma'}\rangle_c
=-\frac{1}{\pi}\int d(\omega')f(\omega')\textrm{Im}
\gl\hat{n}_{\sigma'}f_{\sigma''};f^{\dag}_{\sigma''}\gr_c=-\frac{1}{\pi}\int
d(\omega')f(\omega')\textrm{Im}(\gl\hat{n}_{\sigma'}f_{\sigma''};f^{\dag}_{\sigma''}\gr-\bar{n}_{\sigma'}\gl
f_{\sigma''};f^{\dag}_{\sigma''}\gr)\nonumber\\
&\langle\hat{n}_{\sigma''}c^+_{k\sigma'}f_{\sigma'}\rangle_c=-\frac{1}{\pi}\int
d(\omega')f(\omega')\textrm{Im}
\gl c^{\dag}_{\sigma'}f_{\sigma'}f_{\sigma''};f^{\dag}_{\sigma''}\gr_c\nonumber=-\frac{1}{\pi}\int d(\omega')f(\omega')\textrm{Im}(\gl
c^{\dag}_{\sigma'}f_{\sigma'}f_{\sigma''};f^{\dag}_{\sigma''}\gr-\langle
c^{\dag}_{\sigma'}f_{\sigma'}\rangle\gl
f_{\sigma''};f^{\dag}_{\sigma''}\gr)
\end{split}\end{equation}
where the two-particle Greens function can be obtained from the
single-electron Greens function together with Eq.~\eqref{eq:nff} and Eq.~\eqref{eq:cfff}.
Finally the two connected correlations are
\begin{eqnarray}
\langle \hat{n}_{\sigma''}\hat{n}_{\sigma'}\rangle_c
&=&-\frac{1}{\pi}\int
d(\omega)f(\omega)\textrm{Im}\frac{1}{(N-1)U}\big((\omega-\varepsilon_d-\Delta-(N-1)U\bar{n}_{\sigma'})\gl
f_{\sigma''};f^{\dag}_{\sigma''}\gr\big)\\
\langle\hat{n}_{\sigma''}c^+_{k\sigma'}f_{\sigma'}\rangle_c&=&\frac{-\frac{1}{\pi}\int
d(\omega)f(\omega)\textrm{Im}\Big\{\Big[\frac{-\frac{V_{k\sigma'}}{(N-1)U}(\omega-\varepsilon_d-\Delta)+\Delta\langle
c^{\dag}_{k\sigma'}f_{\sigma'}\rangle+\sum_{k'}V_{k'\sigma'}\langle
c^{\dag}_{k\sigma'}c_{k'\sigma'}\rangle}{\omega-\varepsilon_k-2\varepsilon_d-U-2(N-2)U\bar{n}_{\sigma'}}-\langle
c^{\dag}_{\sigma'}f_{\sigma'}\rangle\Big]\gl
f_{\sigma''};f^{\dag}_{\sigma''}\gr+D\Big\}}{1-\frac{2(N-2)}{\pi}\int
d(\omega)f(\omega)\text{Im}\frac{\gl
f_{\sigma''};f^{\dag}_{\sigma''}\gr}{\omega-\varepsilon_k-2\varepsilon_d-U-2(N-2)U\bar{n}_{\sigma'}}}~~~~
\end{eqnarray}
\end{widetext}
where
\begin{eqnarray}
D=\frac{\frac{V_{k\sigma'}}{(N-1)U}+\langle
c^{\dag}_{\sigma'}f_{\sigma'}\rangle}{\omega-\varepsilon_k-2\varepsilon_d-U-2(N-2)U\bar{n}_{\sigma'}}\label{eq:arbitraryND}
\end{eqnarray}
Compared to Ref.~\onlinecite{lacroix82}, the set of equations
\eqref{eq:arbitraryNGF}-\eqref{eq:Deltatilde} are generalized to
arbitrary degeneracy $N$. Ref.~\onlinecite{czycholl85} has equations
of motion at the same level, but there the three particle Greens functions
are neglected in the limit $U\rightarrow\infty$, while the Greens
functions involving two $c$ operators are considered to give little
contribution for $V\rightarrow0$. Thus, the decoupling of
Ref.~\onlinecite{czycholl85} is constructed in the limit of parameters
$U\rightarrow\infty$, $V\rightarrow0$. In Ref.~\onlinecite{luo99}, the
single impurity Anderson model is studied for infinite interaction
strength $U$ and $N=2$, and the focus is on the approximation beyond
that of Ref.~\onlinecite{lacroix81} with the decoupling scheme of
Ref.~\onlinecite{wang94}.  Here we have implemented the system of equations
\eqref{eq:arbitraryNGF}-\eqref{eq:Deltatilde} for finite $U$ with
arbitrary degeneracy $N$ while neglecting the two particle connected
correlations $\langle \hat{n}_{\sigma''}\hat{n}_{\sigma'}\rangle_c$
and $\langle\hat{n}_{\sigma''}c^+_{k\sigma'}f_{\sigma'}\rangle_c$.

If we now specialize to degeneracy $N=2$ and use the Hermitian
conjugate $\langle f^{\dag}_{\sigma'}c_{k\sigma'}\rangle=\langle
c^+_{\sigma'}f_{\sigma'}\rangle$, the set of equations
\eqref{eq:arbitraryNGF}-\eqref{eq:Deltatilde} becomes
\begin{eqnarray}
\gl
f_{\sigma};f^{\dag}_{\sigma}\gr=\frac{1+\frac{U}{\omega-\varepsilon_d-U-2\Delta-\tilde{\Delta}}\big\{\bar{n}_{\sigma'}+I_1\big\}}{\omega-\varepsilon_d-\Delta-\frac{U}{\omega-\varepsilon_d-U-2\Delta-\tilde{\Delta}}\big\{I_1\cdot\Delta+I_2\big\}}~
\label{eq:N2GF}\end{eqnarray}
with
\begin{eqnarray}
I_1&=&\sum_k\Big(\frac{V_{k\sigma'}\langle
f^{\dag}_{\sigma'}c_{k\sigma'}\rangle}{\omega-\varepsilon_k}-\frac{V^{\ast}_{k\sigma'}\langle
f^{\dag}_{\sigma}c_{k\sigma}\rangle}{\omega+\varepsilon_k-2\varepsilon_d-U}\Big)\label{eq:N2I1}\\
I_2&=&-\sum_{kk'}\Big(\frac{V_{k\sigma'}V^{\ast}_{k'\sigma'}\langle
c^{\dag}_{k'\sigma}c_{k\sigma}\rangle}{\omega-\varepsilon_k}+\frac{V^{\ast}_{k\sigma'}V_{k'\sigma'}\langle
c^{\dag}_{k'\sigma}c_{k\sigma}\rangle}{\omega+\varepsilon_k-2\varepsilon_d-U}\Big)~~~~~~
\label{eq:N2I2}\end{eqnarray}
We calculate the correlations self-consistently from the
spectral theorem
\begin{eqnarray}
~~\langle f^{\dag}_{\sigma}c_{k\sigma}\rangle&=&-\frac{1}{\pi}\int
d\omega' f(\omega'){\rm Im}\frac{V^{\ast}_{k\sigma'}\gl
f_{\sigma};f^{\dag}_{\sigma}\gr}{\omega'-\varepsilon_k}\\
\langle c^{\dag}_{k'\sigma}c_{k\sigma}\rangle&=&-\frac{1}{\pi}\int
d\omega'
f(\omega'){\rm Im}\gl c^{\dag}_{k\sigma};c_{k'\sigma}\gr\nonumber\\
&=&-\frac{1}{\pi}\int d\omega' f(\omega'){\rm
Im}\Big\{\frac{\delta_{kk'}}{\omega-\varepsilon_{k'}}\nonumber\\
&&\qquad\qquad\qquad+\frac{V^{\ast}_{k\sigma'}V_{k'\sigma'}\gl
f_{\sigma};f^{\dag}_{\sigma}\gr}{(\omega-\varepsilon_k)(\omega-\varepsilon_{k'})}\Big\}~~~~~
\end{eqnarray}
where $f(\omega')$ is the Fermi distribution function and the equation
of motion for $\gl c^{\dag}_{k\sigma};c_{k'\sigma}\gr$ has been used.

\section{Methods of solution}
\label{sec:ga}

In principle, the system of equations \eqref{eq:N2GF}-\eqref{eq:N2I2}
can be solved iteratively. But it turns out that the iterative
solution requires significant Lorentzian broadening $\omega\to
\omega+i\eta$ and very small linear mixing factors $\alpha$.
Furthermore, there are parameter regimes for which it is hard to
converge a solution. The situation is not significantly improved by
better mixing schemes like Broyden mixing~\cite{srivastava84}.
Therefore, we turned to a different approach for finding the
selfconsistent solutions. Genetic algorithms (GA) are adaptive
heuristic search algorithms based on the idea of evolution by natural
selection~\cite{goldberg89,rutkowski08} and have been used in many
optimization or minimization problems of science and
engineering~\cite{gen08}. In
Refs.~\onlinecite{grigorenko00,grigorenko01}, the GA method was
employed to calculate the ground state wave function of one- and
twodimensional quantum systems. We adopt this idea of optimizing the
wave function until they obey the Schr\"odinger equation and carry it
over to our optimization problem, that of finding a Greens function
$G(\omega)=\gl f_{\sigma};f^{\dag}_{\sigma}\gr$ that fulfils
Eq.~\eqref{eq:N2GF}. This approach turns out to significantly improve
the convergence speed, and as will be demonstrated below, the fact
that it works with little or no broadening, the solutions are
qualitatively better than from an iterative approach.  The increase in
convergence speed is essential for application of the SIAM model
solution in DMFT calculations.

The GA algorithm is started with a ``population'' of initial guesses.
The imaginary parts of the initial population of trial Greens
functions are guessed as sums of Gaussians
\begin{eqnarray}
{\rm Im}\,G(\omega)=L\Big(e^{-\frac{(\omega-B)^2}{2C^2}}+e^{-\frac{(\omega-B-U)^2}{2C^2}}\Big)
\end{eqnarray}
where $L$ is a normalization factor, $B,C$ are randomly generated
numbers and $U$ is Coulomb interaction strength. We use the
Kramers-Kronig relation to determine the real part
\begin{eqnarray}
{\rm Re}\,G(\omega)=-\frac{1}{\pi}\int
\frac{{\rm Im}\,G(\omega)}{\omega'-\omega}d\omega'
\end{eqnarray}
The convergence of the method can be speeded up if the positions of
the randomly generated peaks cumulate around the positions of the
Hubbard bands known from the atomic limit.  Besides Gaussians, we have
also tested other functional forms of the initial guess, but this had
little influence on convergence speed and final converged result.

The generation of trial Greens functions is evaluated and ordered
according to a ``fitness function'' which measures the closeness to a
selfconsistent solution.  Thus, we define the fitness function as
\begin{eqnarray}
F[G(\omega)]=\|G(\omega)-{\rm rhs}[G(\omega)]\|
\end{eqnarray}
where ${\rm rhs}[G(\omega)]$ represents the right hand sides of
Eq.~\eqref{eq:arbitraryNGF} or Eq.~\eqref{eq:N2GF} which are
functionals of $G(\omega)$ via the integral terms
\eqref{eq:arbitraryNA}-\eqref{eq:arbitraryNC} or
\eqref{eq:N2I1}-\eqref{eq:N2I2}.  The norm
\begin{equation}
\|f(\omega)\| = \int d\omega |f(\omega)|
\end{equation}
measures the distance of the trial Greens functions from the
selfconsistent solutions of Eqs.~\eqref{eq:arbitraryNGF} or
\eqref{eq:N2GF}.

\begin{figure}[tbp]
\centering
\includegraphics[width=0.45\textwidth]{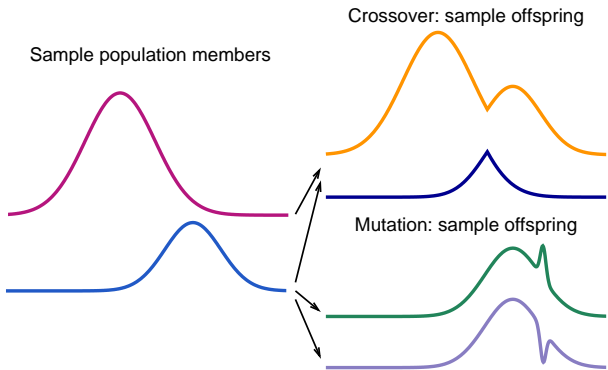}
\caption{(Color Online) Illustration for the genetic operations of crossover and mutation. The shown curves represent imaginary parts of candidates for $G(\omega)$. }\label{fig:crossover}
\end{figure}

According to the standard procedure of GA, a new ``generation'' of the
population of trial Greens functions is formed by application of two
GA operators, ``crossover'' and ``mutation''.  The crossover operation
is
\begin{equation}\begin{split}
{\rm Im}\,G_1^{\rm offspring}(\omega)=L_1{\rm Im}\,\Big\{&G^{parent}_1(\omega)\Theta(\omega-\omega_0)\\&+G^{parent}_2(\omega)\Theta(\omega_0-\omega)\Big\}\\
{\rm Im}\,G_2^{\rm offspring}(\omega)=L_2{\rm Im}\,\Big\{&G^{parent}_1(\omega)\Theta(\omega_0-\omega)\\&+G^{parent}_2(\omega)\Theta(\omega-\omega_0)\Big\}
\end{split}\end{equation}
where $\omega_0$ is the randomly chosen crossover position,
$\Theta(\omega)$ is the Heaviside function and $L_1$, $L_2$ are
normalization factors. Mutation introduces, with a low probability,
random small changes in the trial Greens function in order to prevent
the population from stabilizing in a local minimum. The mutation
operator is
\begin{eqnarray}
\textrm{Im}\,G^{\rm offspring}(\omega)=L\big(\textrm{Im}\,G^{parent}(\omega)+
Ae^{-\frac{(\omega-B)^2}{2C^2}}\big)
\end{eqnarray}
where $A, B, C$ are randomly generated numbers and $L$ normalizes the
function.  For both crossover and mutation, real parts are obtained
via the Kramers-Kronig relation. Crossover and mutation operations are
illustrated in Fig.~\ref{fig:crossover}.

Now the principles of selection have to be discussed. Some of the best
members of a generation are preserved without replacing them by their
offspring.  Furthermore, the trial Greens functions that are actually
included into the new generation are obtained by entering the result of
the GA operations into Eqs.~\eqref{eq:arbitraryNGF} or \eqref{eq:N2GF}
and calculating one iterative step. While in principle, pure GA
operations could be used to find an optimal solution, the strict
requirements imposed on a selfconsistent solution are more easily met
by alternation of GA operations and iterative steps.  As the new
generation has more than twice as many members as the previous one,
many are dropped according to their fitness values.  In order to avoid
premature convergence of the population to a suboptimal solution, some
members with unfavorable fitness values are kept in the population,
and some new random trial Greens functions are added to the
population.  The end of the evolution is determined, as in the
iterative solution of the integral equations, by a member of the
population reaching the target accuracy. We usually use fitness
function values of $10^{-3}$ as a criterion for terminating the GA
procedure.  An additional advantage of the GA approach is the ease
with which it can handle arbitrary kinds of constraints; they can be
included as weighted components of the fitness function.

\section{Results and discussions}
\label{sec:results}

\begin{figure}[tbp]
\centering
\includegraphics[angle=-90,width=0.45\textwidth]{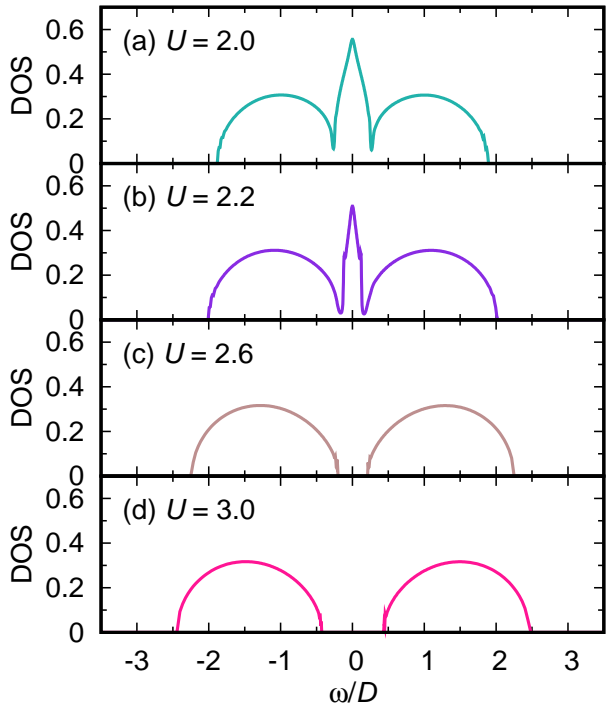}
\caption{(Color Online)  DOS calculated with GA method for particle-hole
symmetric Hubbard model on Bethe lattice}\label{fig:GA_Hubbard}
\end{figure}

First, we investigated band-width control Mott metal-insulator
transition in the Hubbard model. The densities of states (DOS) at
four different values of $U$ are shown in Fig.~\ref{fig:GA_Hubbard}.
As expected, quasi-particle peak as well as the upper and lower
Hubbard bands are present in the metallic phase and transfer of
spectral weight from quasi-particle peak to the Hubbard bands is
clearly evident by reduction of the width of the central peak. In
the insulating state, the central peak suddenly vanishes and a gap
appears between upper and lower Hubbard bands. Further increasing
$U$ leads to an increasing gap amplitude. The critical value of $U$
for Mott transition obtained from our impurity solver is
$U_c\approx2.5$. Compared to the critical value from numerical
renormalization group method where $U_c\approx2.94$~\cite{Bulla2},
our result underestimates the critical value of U due to the
decoupling scheme. We note that in the metallic region, the height
of our obtained DOS at the Fermi level is not fixed. This is due to
the fact that two peaks in the imaginary part of the self-energy are
quite close to the Fermi level, resulting in a numerical difficulty
in getting a vanishing value of the imaginary part of the
self-energy at Fermi level.

\begin{figure}[tbp]
\centering
\includegraphics[angle=-90,width=0.45\textwidth]{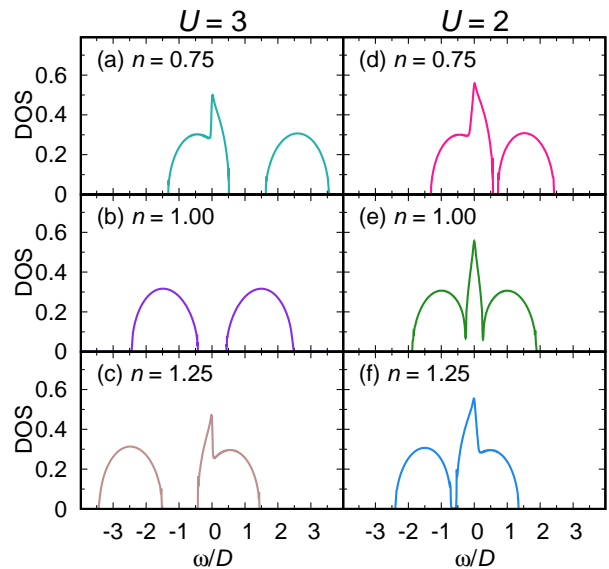}
\caption{(Color Online)  DOS for the asymmetrical Hubbard model on the Bethe
lattice: filling controlled metal insulator transition.}\label{fig:filling}
\end{figure}

\begin{figure}[tbp]
\centering
\includegraphics[angle=-90,width=0.45\textwidth]{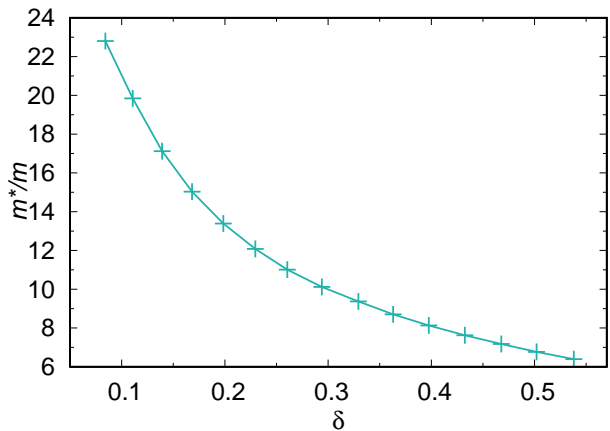}
\caption{(Color Online)  Effective mass at different fillings for the Hubbard
model on the Bethe lattice.}\label{fig:mass}
\end{figure}

\begin{figure}[tbp]
\centering
\includegraphics[angle=-90,width=0.45\textwidth]{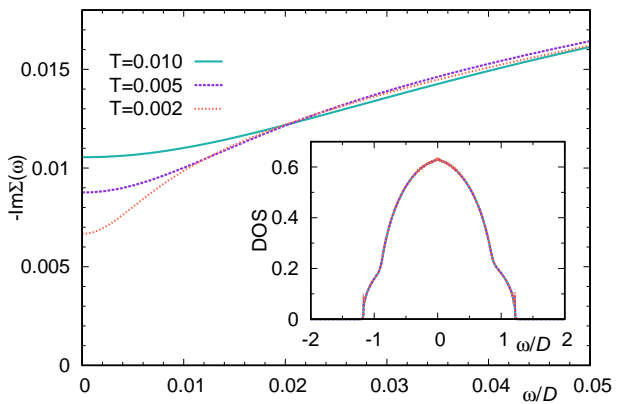}
\caption{(Color Online) Self-energy at low temperature close to the
  Fermi level for the metallic state ($U=1$) for the particle-hole
  symmetric Hubbard model on the Bethe lattice. The inset shows the
  corresponding DOS. }\label{fig:fermiliquid}
\end{figure}

Then, let us study the filling controlled Mott metal-insulator
transition on the Hubbard model. In Fig.~\ref{fig:filling}, we
present the DOS as a function of doping at two different values of
$U$. It is found that filling controlled metal-insulator transition
occurs at $U=3$ while the system remains in metallic state at $U=2$.
At $U=3$, we also investigate the effective mass
\begin{eqnarray} \frac{m^{\ast}}{m}=1-\frac{\partial{\rm
Re}\Sigma(\omega)}{\partial\omega}\Big{|}_{\omega\rightarrow 0}
\end{eqnarray}
as a function of doping concentration. It is shown in
Fig.~\ref{fig:mass} that the effective mass clearly displays a
divergent behavior as doping concentration goes to zero which seems to
obey Brinkman-Rice picture for the Fermi liquid~\cite{Brinkman70}. In
the small doping region, the carriers are more easily localized. We
also studied the low frequency behavior near the Fermi level for the
metallic state at different temperatures, as shown in
Fig.~\ref{fig:fermiliquid}. We obtained that the imaginary part of the
self-energy does not exactly follow Fermi liquid behavior under the
present decoupling scheme. However, as the temperature approaches
zero, the negative imaginary part of the self-energy decreases. The
precision of the results at very low temperature is presently limited
numerically. Therefore, the exact behavior of the imaginary part of
the self-energy at the Fermi level at zero temperature is beyond our
reach.  Even though our decoupling scheme qualitatively shows an
acceptable behavior, from principal considerations exact Fermi liquid
behavior is not to be expected from a decoupling approach.

\begin{figure}[tbp]
\centering
\includegraphics[angle=-90,width=0.45\textwidth]{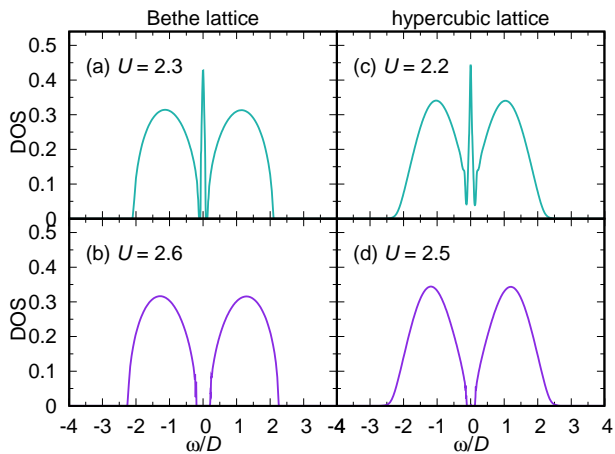}
\caption{(Color Online) Comparison of Hubbard model with different kinds
of bath: (left) Semicircle bath on Bethe lattice (right) Gaussian
bath on hypercubic lattice. top two figures are metallic state,
bottom two figures are insulator state just away from transition
point}\label{fig:different_bath}
\end{figure}

We have also studied the Hubbard model with different types of bath
as shown in Fig.~\ref{fig:different_bath}. The influence of the bath
has often been considered to be small since the self-consistent
solution will not depend much on the initial guess of the bath. Our
result shows that both Bethe lattice and hypercubic lattice produce
qualitatively similar results for the Mott transition. However, our
results show that different baths yield different critical
interactions strengths $U_c$ at which the Mott transition sets in.
For the Bethe lattice, we find $U_c^{\rm Bethe}\approx 2.5$, while
for the hypercubic lattice, the result is $U_c^{\rm
hypercubic}\approx2.4$.

\begin{figure}[tbp]
\centering
\includegraphics[angle=-90,width=0.45\textwidth]{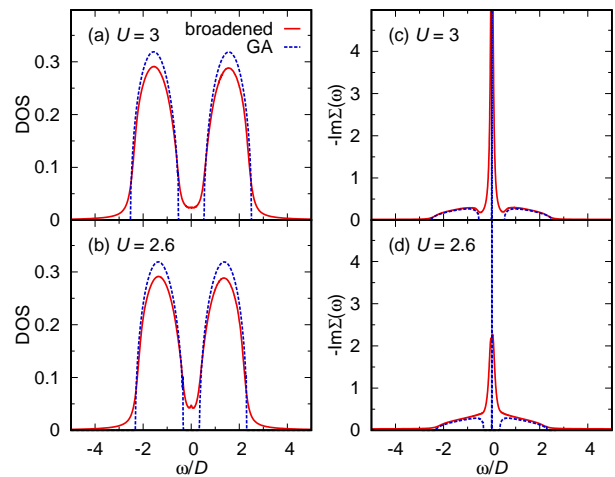}
\caption{(Color Online) Comparison between the GA method and the iterative method
with Lorentzian broadening for the particle-hole symmetric Hubbard
model on the Bethe lattice: (top) DOS and self-energy for the
insulating state with $U=3$. (bottom) DOS and self-energy close to
the Mott transition with $U=2.6$.}\label{fig:comparison}
\end{figure}

We now turn to a comparison of the two methods of solution we
employed, the iterative method with Lorentzian broadening and the
combined genetic algorithm and iterative method. The comparison of
the CPU time needed for a selfconsistent solution is clearly in
favor of GA: While in general the solution with pure iteration takes
four times longer, near the Mott transition the iterative solution
becomes very slow and inefficient.  In Fig.~\ref{fig:comparison}, we
show the results obtained with both methods for the same parameter
values.  In Fig.~\ref{fig:comparison}, top left, we can see that in
the DOS there is a nonzero continuous connection between two Hubbard
bands in the result obtained with Lorentzian broadening, which makes
it difficult to distinguish the Mott transition clearly when $U$
approaches the critical value for the transition $U_c$ because the
quasiparticle peak is very small in that case. Moreover, the Kondo
peak will be greatly influenced by the amount of broadening.
Different broadening will give different critical value of $U_c$.
However, the combined GA and iteration method can give more precise
results near the critical point which can be seen from the bottom
left DOS figure. For the combined GA method, at $U=2.6$ we find an
insulating state, while the Lorentzian broadening method still gives
a metallic state at Fermi surface. This is due to the fact that the
divergent behavior of the imaginary part of the self-energy just
above the Mott transition cannot be correctly captured if there
exists a finite broadening factor. However, in the GA method, the
broadening factor can be even set to zero, which eliminates the
numerical problem induced by the factor. The right panels of
Fig.~\ref{fig:comparison} shows the comparison of the imaginary part
of the self-energy. It is found that GA method really gives a
correct divergent behavior even close to the Mott transition at the
Fermi level, while the Lorentzian broadening method does less well.

\begin{figure}[htbp]
\centering
\includegraphics[angle=-90,width=0.45\textwidth]{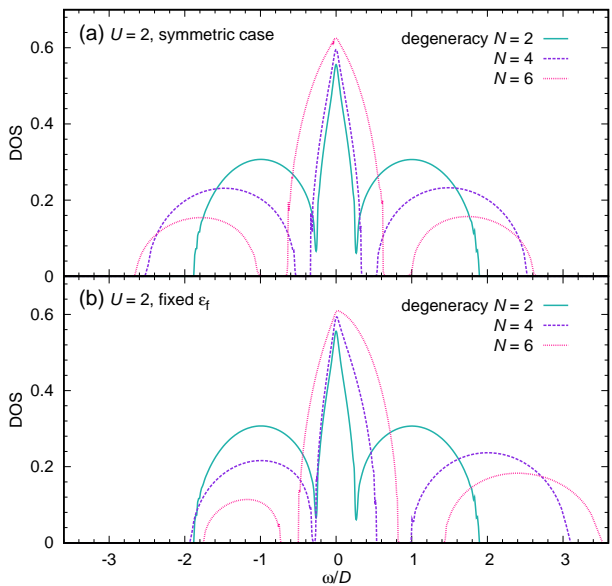}
\caption{DOS for the finite $U$ Hubbard model  for
several values of the spin-orbital degeneracy $N$. The interaction
strength is $U=2$, the hybridization strength $V=0.25$. (a) half-filling, (b) fixed impurity position $\varepsilon_f$. }\label{fig:arbitraryN}
\end{figure}

We have also studied the Hubbard model with arbitrary degeneracy $N$.
We have found that the decoupling scheme works nicely for $N=2$, but
for $N>2$ there are some deviations from particle hole symmetry at
half filling. We observe that the band positions and occupation
numbers are correct, but some broadening of the upper Hubbard band is
missing.  This shows that the presently used decoupling of the
three-particle Greens functions misses some terms that would
contribute to the damping of the upper Hubbard band. We remedy this
small deficiency by adding the {\it ansatz} $S(N)=-c
(N-2)\Delta(\omega)$ in the denominator of $A$ (see
Eq.~\ref{eq:arbitraryNA}) because this denominator is mainly
responsible for the upper Hubbard band, and neglected contributions
from higher order Greens functions should contribute an unknown
function of $\Delta(\omega)$. The factor $c=0.5$ is found numerically
from the requirement of particle hole symmetry at half filling, and
the correction acts only for $N>2$ as $S(2)\equiv 0$.  We get the
results shown in Fig.~\ref{fig:arbitraryN}, where we have calculated
the spectral functions for various degeneracies $N$ at a temperature
$T=0.01$. With increasing $N$, the total on-site Coulomb interaction
increases so that the two Hubbard bands shift further away from the
Fermi level. But the critical $U_c$ also increases with $N$.
Therefore, at the same $U$, the system shows more metallicity for
larger $N$, and transfer of spectral weight is observed from upper and
lower Hubbard bands to the Kondo peak with increasing $N$.
This result is consistent with the QMC result of
Ref.~\onlinecite{Han98}.

\begin{figure}[tbp]
\centering
\includegraphics[angle=-90,width=0.45\textwidth]{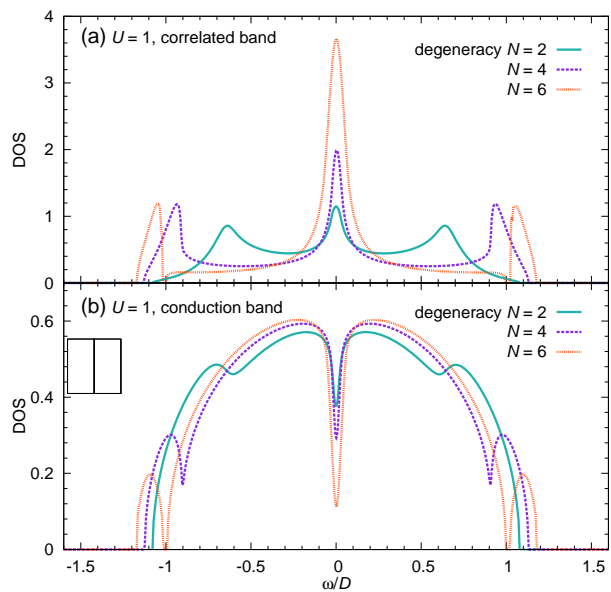}
\caption{(Color Online)  DOS for the periodic Anderson model for different values
  of the spin-orbital degeneracy $N$. (a) correlated band, (b)
  conduction band.}\label{fig:PAM}
\end{figure}

For comparison, we have calculated the Periodic Anderson model with
our code in Fig.~\ref{fig:PAM}. We observe a similar behavior of
the spectral weight transfer as in the large $N$ Hubbard model.
Meanwhile, Fig.~\ref{fig:PAM} differs from the behavior shown in
Ref.~\onlinecite{qi08}.

\section{Conclusions}

We have presented the derivation and implementation of a solution of
the single impurity Anderson model based on equations of motion and
truncation. We employ a combination of genetic algorithms and
iteration to solve the resulting integral equations. We demonstrate
that our method is useful as an impurity solver in the context of
dynamical mean field theory. We show results for the Mott metal
insulator transition as a function of interaction strength $U$ and
as a function of filling $n$, and also show the trend of weight
transfer at different values of the spin-orbital degeneracy $N$.

We gratefully acknowledge support of the DFG through the Emmy Noether program.

\end{document}